\documentclass[aip,cha,onecolumn,superscriptaddress,preprintnumbers,showpacs, amsmath,amssymb]{revtex4-1}
\usepackage{bm}
\usepackage[colorlinks=true,linkcolor=blue,citecolor=blue]{hyperref}
\usepackage{times}
\usepackage{amsmath}
\usepackage{amssymb}
\usepackage{amsthm}
\usepackage{amsfonts}
\usepackage{enumerate}
\usepackage{latexsym}
\usepackage{ifpdf}
\usepackage{graphicx}
\usepackage{makeidx}
\expandafter\ifx\csname package@font\endcsname\relax\else
 \expandafter\expandafter
 \expandafter\usepackage
 \expandafter\expandafter
 \expandafter{\csname package@font\endcsname}%
 \fi
\usepackage{color}

\begin{document}

\title{Metallic interface in non-SrTiO$_3$ based titanate superlattice}
\author{Xiaoran Liu}
\email{xxl030@uark.edu}
\affiliation{Department of Physics, University of Arkansas, Fayetteville, Arkansas 72701, USA}

\author{D. Choudhury}
\affiliation{Department of Physics, University of Arkansas, Fayetteville, Arkansas 72701, USA}
\affiliation{Department of Physics, Indian Institute of Technology, Kharagpur 721302, INDIA}

\author{Yanwei Cao}
\affiliation{Department of Physics, University of Arkansas, Fayetteville, Arkansas 72701, USA}

\author{M. Kareev} 
\affiliation{Department of Physics, University of Arkansas, Fayetteville, Arkansas 72701, USA}

\author{S. Middey}
\affiliation{Department of Physics, University of Arkansas, Fayetteville, Arkansas 72701, USA}

\author{J. Chakhalian}
\affiliation{Department of Physics, University of Arkansas, Fayetteville, Arkansas 72701, USA}

\begin{abstract}

We report on the fabrication of all perovskite Mott insulator/band insulator YTiO$_3$/CaTiO$_3$ superlattices by pulsed laser deposition. The combination of {\it{in-situ}} reflection high energy electron diffraction, X-ray diffraction and X-ray reflectivity confirms the high quality of the films grown in a layer-by-layer mode. Electrical transport measurements reveal that a non-SrTiO$_3$ based two-dimensional electron gas system has formed at the YTiO$_3$/CaTiO$_3$ interface. These studies offer another route in the pursuit of complex oxide two-dimensional electron gas systems which allows to obtain greater insights into the exotic many-body phenomena at such interfaces. 

\end{abstract}

\maketitle

\newpage

In recent years two-dimensional electron gases (2DEGs) at the interface of complex insulating oxide heterostructures have attracted tremendous research interest because of the observed emergent physical phenomena as well as the potential device applications.\cite{Allen, Bristowe, Nakagawa, Zubko, Igor, Mannhart} While a large number of perovskite based heterostructures have been discovered experimentally so far\cite{Hwang,Perna,Hotta,Hwang2002,Santosh,Pouya,Dai,Zou}, all of these systems contain layers of quantum paraelectric SrTiO$_3$ (STO) \cite{Muller} except for the notable LaTiO$_3$/KTaO$_3$\cite{Zou}. Among these STO-based systems, specifically for the $R$TiO$_3$/SrTiO$_3$ ($R$ represents a trivalent rare-earth ion) family, it was pointed out that the interfacial electronic properties and magnetism can be effectively tailored by the specific choose of $R$.\cite{Santosh,Moetakef_PRX,Jackson,Jang} Based on this, it can be conjectured that similar behavior can be achieved by substituting Sr with a divalent \textit{alkaline} earth ion, $A$ in $R$TiO$_3$/$A$TiO$_3$ superlattices\cite{Ganguli}. Up to date, however, there is no experimental information on whether any $R$TiO$_3$/$A$TiO$_3$ interface outside of STO support metallic ground state. 

In this Letter, we report on the fabrication of ultra-thin perovskite 3YTiO$_3$/6CaTiO$_3$ (3 and 6 refers to the pseudocubic unit cell) superlattice on NdGaO$_3$ (110) substrate, which exhibits a 2DEG \textit{without} STO. The schematic structure of the superlattice is displayed in Fig. 1(a). In the bulk, orthorhombic YTiO$_3$ (YTO) is a 3$d$$^1$ Mott insulator (Mott gap $\sim$ 1.2 eV\cite{Okimoto}) that undergoes the para- to ferro- magnetic phase transition at 30 K\cite{Chae} while orthorhombic CaTiO$_3$ (CTO) is a 3$d$$^0$ diamagnetic band insulator with an energy gap of $\sim$ 3.5 eV\cite{Ali, Ueda}. (110)-oriented orthorhombic NdGaO$_3$ (NGO) is selected as the substrate because it offers the symmetry continuity and small lattice mismatch with the film, as shown in Table I. The entire growth process was monitored by {\it{in-situ}} reflection high energy electron diffraction (RHEED) demonstrating good crystallinity and smooth surface morphology. X-ray diffraction (XRD) and X-ray reflectivity (XRR) establish the epitaxial growth of the superlattice with abrupt interface. Transport measurements reveal the superlattices exhibit metallic conduction from 300 K down to 2 K with electrons as the carriers.        


A set of 3YTO/6CTO superlattices was fabricated by pulsed laser deposition using a KrF excimer laser operating at $\lambda$  $=$ 248 nm. During  growth, the NGO substrate was kept at 750 $^{\circ}$C in a high-vacuum atmosphere of 10$^{-6}$ Torr to avoid the formation of Y$_2$Ti$_2$O$_7$ impurity phase\cite{Chae}. The laser fluency and pulse repetition rate were set at about 1.4 J/cm$^2$ and 2 Hz, respectively. After growth samples were cooled down to room temperature without annealing at a rate of 15 $^{\circ}$C/min. Structural properties of the sample were characterized by X-ray Diffraction (XRD) and X-ray Reflectivity (XRR) measurements (Cu K$_{\alpha1}$ line, $\lambda$ = 1.5406 {\AA}). The electrical transport properties were performed by a Physical Property Measurement System (PPMS, Quantum Design) using the Van der Pauw geometry. The Hall effect measurements were carried out in an external magnetic field of up to 7 T oriented normal to the sample surface.


Figure 1(b) shows the RHEED pattern of the NGO (110) substrate before growth. As seen, the characteristic specular (0 0) and off-specular (0 $\pm$1) Bragg reflections on the first Laue circle testify for the smooth morphology of the substrate. The half orders indicated by the white arrows are due to the orthorhombic (001) reflections.\cite{Proffit} During the alternating deposition of YTO and CTO layers, the distinct spots from specular and off-specular reflections with the characteristic streak pattern shown in Fig. 1(c) and 1(d) were recorded and confirm the good crystallinity and flatness of the superlattice. Importantly, the half order reflections were still observed after cooling the sample down to room temperature, indicating that the superlattice preserves the orthorhombic symmetry.  

In addition to morphological quality, it is of critical importance to ensure that the superlattices contain no  impurity phase. To establish this we performed the XRD $\omega$-2$\theta$ scans in the vicinity of the NGO (002) reflection on a [3YTO/6CTO]$_7$ sample. As seen in Fig. 2(a), besides the film (002) reflection at 23.826$^{\circ}$ on the right side of the sharp substrate peak, the first order superlattice peak is also observed at $\sim$ 22.389$^{\circ}$. This yields a thickness of 33.39 {\AA} for one repeat (3YTO + 6CTO) that is very close to the expected value of 34.35 {\AA}. The total thickness of the sample estimated from the Kiessig fringes is  25.24 nm, which agrees very well with the expected value. Complimentary detailed information about the superlattice thickness as well as the surface and interface roughness was obtained from XRR measurements; a model fitting of the experimental reflectivity data yields the film thickness of $\sim$ 25.16 nm which is in excellent agreement with our XRD results. The obtained average interface and surface roughness are 3 {\AA} and 5 {\AA}, respectively. These results further confirm that all grown superlattices have good crystallinity with the sharp interface and surface. 

Having confirmed the high structural quality of the samples, we turn to explore the electrical transport properties of the superlattice. As seen in Fig. 3(a), the resistance of the superlattice (purple curve) exhibits distinct metallic behavior in the whole temperature range, i.e. it decreases continuously as the temperature is lowered down to the base temperature. The inset in Fig. 3(a) shows the corresponding temperature-dependent sheet resistance per YTO/CTO interface. In order to assure that the metallic behavior is due to the interface and elucidate the possible role of defect or oxygen doping, we performed the same measurements on two reference samples: 22 u.c. thick CTO on NGO and 22 u.c. thick YTO on NGO, both fabricated under the identical growth conditions as the superlattices. As immediately seen in Fig. 3(a), the 22 u.c. CTO film shows strongly insulating behavior (black curve) until its resistance goes out of the range around 120 K; at the same time the resistance of YTO is beyond the measurement range of our apparatus even at room temperature. By comparing the superlattices with these reference samples it allows to confirm that the observed metallic conductivity of the superlattices indeed emerges from the YTO/CTO interface. In addition, it is worth noting that YTO/STO superlattice was reported to be insulating\cite{Misha} which corroborates the essence of the combination of YTO and CTO.

To establish the type of charge carries and estimate the carrier density we performed the Hall resistivity measurements. The obtained Hall resistance data exhibits linear relationship with the applied magnetic field and shows a negative slope, $R_H$ implying that electron is the type of charge carriers in the superlattices. This strongly indicates the interfacial charge transfer process from YTO to CTO and the conducting electrons likely residing on the CTO side, which is consistent with a recent theoretical calculation considering band alignments between various oxides including YTO and CTO.\cite{Bjaalie} The calculated sheet carrier density $n_s$ as a function of temperature is shown in Fig. 3(b). As seen, the magnitude is $\sim$ 10$^{14}$ cm$^{-2}$ and is practically constant with gradual increase towards lower temperatures. The obtained value is quite similar to the previously reported 2DEG systems.\cite{Perna,Dai,Zou,Moetakef}. Further, the corresponding Hall mobility $\mu_H$ is shown in Fig. 3(b) as a function of temperature. As seen, at 300K $\mu$$_H$ has the same magnitude as other STO-based systems but at 10 K $\mu$$_H$ sharply  decreases to $\sim$ 2 cm$^2$V$^{-1}$s$^{-1}$ , which is about 2 order of magnitude smaller than in STO-based 2DEGs.\cite{Perna, Dai, Moetakef}. The origin of reduced mobility could be connected to several scattering mechanisms, which dominate at different temperature: ionic impurity scattering (low temperature), electron-electron scattering (intermediate) and longitudinal optical phonon scattering (high temperature).\cite{Mikheev,Verma} The strong suppression of $\mu_H$ at low temperature in YTO/CTO is possibly due to the combination of scattering from defects present at the interface and large crystal field distortion at the YTO/CTO interface which should give rise to a strongly enhanced ionic impurity scattering.


In summary, we have fabricated the non-STO based ultra-thin 3YTO/6CTO superlattices on NGO (110) substrate exhibiting 2DEG behavior. A combination of RHEED, XRD and XRR measurements confirms the quality layer-by-layer growth of the films with smooth surface morphology and sharp interfaces. Transport measurements reveal that the superlattices exhibit metallic conduction from 300 K to 2 K with electrons as the charge carriers. Our findings suggest another route to realizing complex oxide 2DEGs without hindering effects from complex behavior of STO and provides a system to investigate possible interplay between magnetism and superconductivity at the interface\cite{JC}.


J.C. is supported by the Gordon and Betty Moore Foundation's EPiQS Initiative through Grant GBMF4534. S.M. is supported by the DOD-ARO under Grant No. 0402-17291. Y.C. and X.L. acknowledge the support by the Department of Energy grant DE-SC0012375.

\newpage

\begin{table}[t]
\caption{Lattice parameters of bulk CTO, YTO and NGO. The $a$, $b$ and $c$ represent the values of the orthorhombic unit cell while the $a$$_c$, $b$$_c$ and $c$$_c$ for pseudocubic unit cell.}
\centering
\setlength{\tabcolsep}{12pt}
\begin{tabular}{c c c c c c}
\hline\hline
Material & $a$ (\AA) & $b$ (\AA) & $c$ (\AA) & $a$$_c$ = $b$$_c$ = $\sqrt{a^2+b^2}/2$ (\AA) & $c$$_c$ = $c/2$ (\AA) \\ [0.5ex]
\hline
CTO & 5.380 & 5.442 & 7.640 & 3.826 & 3.820 \\
YTO & 5.316 & 5.679 & 7.611 & 3.889 & 3.806 \\
NGO & 5.428 & 5.498 & 7.708 & 3.863 & 3.854 \\
\hline\hline
\end{tabular}
\label{Table I}
\end{table}
\clearpage

\newpage

\begin{figure*}[t]\vspace{-0pt}
\center
\includegraphics[width=0.45\textwidth]{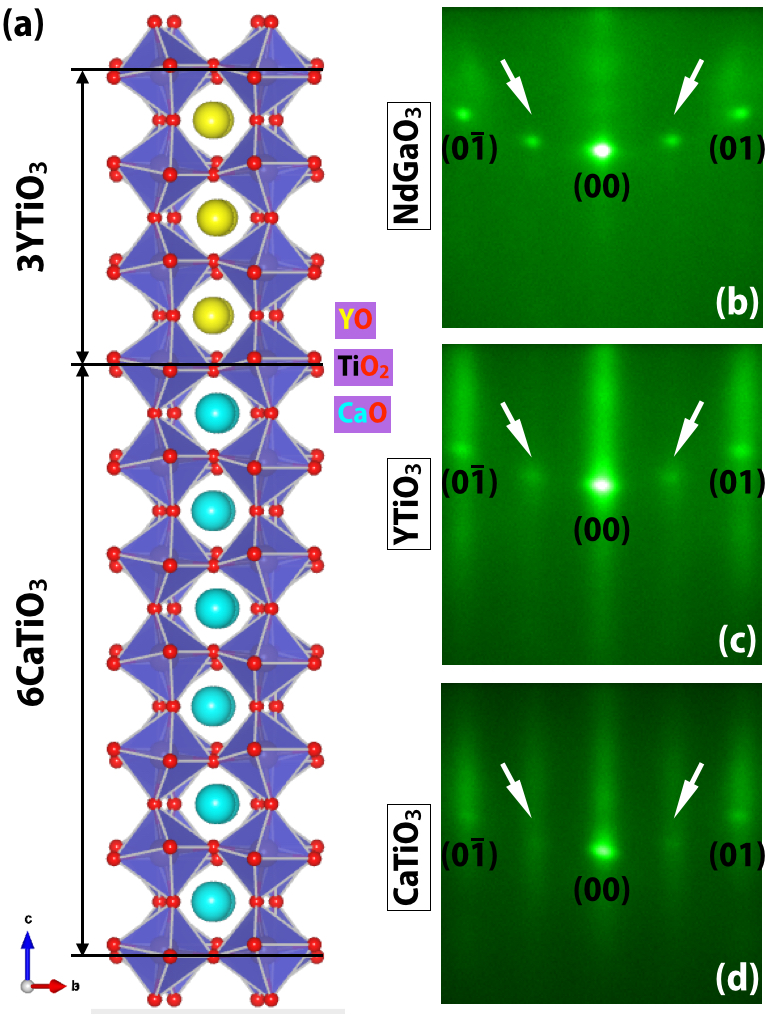}
\caption{\label{} (a) Schematic of the 3YTO/6CTO interface along the [001] growth direction. (b)-(d) $In$-$situ$ RHEED patterns of (b) NGO (110) substrate, (c) YTO layers and (d) CTO layers. The observed half orders indicated by white arrows are from the orthorhombic (001) reflections.}
\end{figure*}

\newpage

\begin{figure*}[t]\vspace{-0pt}
\center
\includegraphics[width=0.5\textwidth]{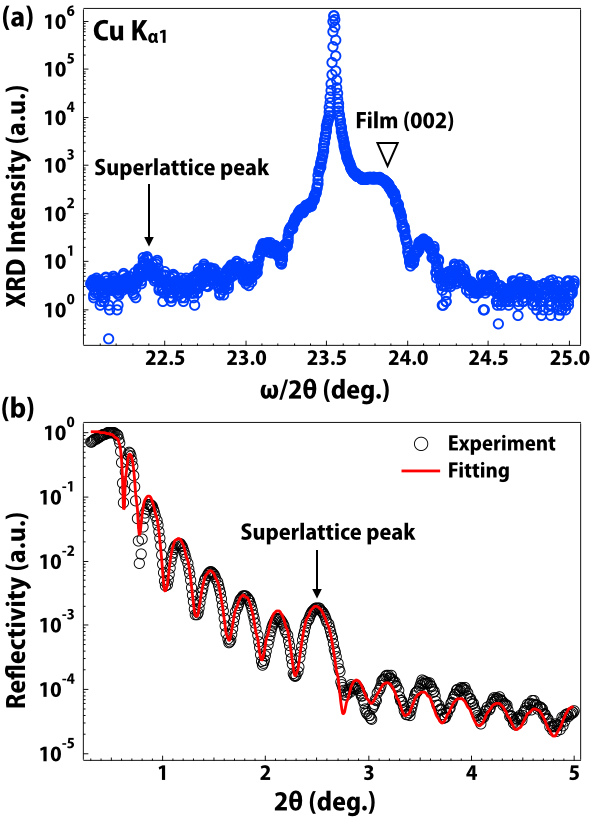}
\caption{\label{} (a) X-ray diffraction $\omega$-2$\theta$ scan of a [3YTO/6CTO]$_7$ superlattice. The film (002) peak and the superlattice satellite peak are labeled on the graph, respectively. (b) X-ray reflectivity data of the same sample. Total thickness of the superlattice calculated according to the Kiessig fringes is about 24 nm.}
\end{figure*}

\newpage

\begin{figure*}[t]\vspace{-0pt}
\includegraphics[width=0.45\textwidth]{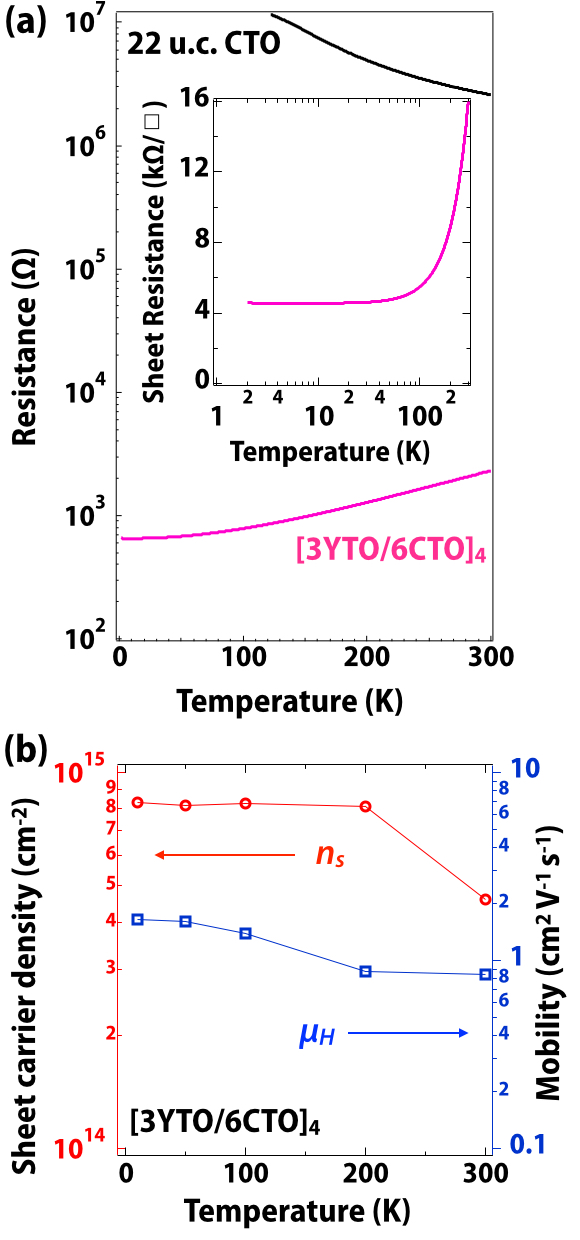}
\caption{\label{} (a) Resistance of the 3YTO/6CTO superlattice (purple curve) and the 22 unit cell CTO film (black curve) vs. temperature. Inset: Sheet resistance of the superlattice vs. temperature in the log scale. Note, temperature is plotted in the log scale to show the absense of upturn behavior down to 2 K. (b) Sheet carrier density and mobility of 3YTO/6CTO.}
\end{figure*}

\end{document}